\documentclass[12pt,osajnl2,preprint,showpacs,superscriptaddress]{revtex4} 
\usepackage[draft]{hyperref} 

\usepackage{amsmath,amssymb,amsfonts}
\usepackage[english]{babel}
\usepackage{graphicx,color}

\begin{document}

\title{On optical Weber waves and Weber-Gauss beams.}

\author{B. M. Rodr\'{\i}guez-Lara}
\affiliation{Institute of Photonic Technologies, National Tsing-Hua University, Hsinchu 300, Taiwan.}
\email{bmlara@mx.nthu.edu.tw}

\begin{abstract}
The normalization of energy divergent Weber waves and finite energy Weber-Gauss beams is reported. The well-known Bessel and Mathieu waves are used to derive the integral relations between circular, elliptic, and parabolic waves and to present the Bessel and Mathieu wave  decomposition of the Weber waves. The efficiency to approximate a Weber-Gauss beam as a finite superposition of Bessel-Gauss beams is also given.
\end{abstract}


\maketitle

\section{Introduction}

Ideal propagation invariant scalar waves are the separable solution to the Helmholtz equation with cylindrical symmetry. There exist four fundamental families related to the four cylindrical coordinate systems for which the reduced wave  equation is separable: Plane waves for Cartesian symmetry, Bessel, Mathieu and Weber waves for circular-, elliptic- and parabolic-cylindrical symmetries, in that order   \cite{Whittaker1927, Stratton1941, Morse1953v1, Miller1984}. 
In optics, the interest in these families of structured scalar waves nowadays lies beyond their propagation invariance \cite{Durnin1987p651,Durnin1987p1499}. It is common knowledge that they all carry a well defined linear momentum in the propagation direction. In addition, Plane waves carry a well defined linear momentum in the $x$ and $y$ directions, Bessel waves carry a well defined orbital angular momentum, Mathieu waves carry a well defined composition of orbital angular and $x$-linear momenta, while Weber waves carry a composition of orbital angular and $y$-linear momenta \cite{Makarov1967p4,Boyer1976p35}. These dynamical properties are acquired by the corresponding beams and vector fields, both in the classical and quantum regime  \cite{Jauregui2005p033411,VolkeSepulveda2006p867,RodriguezLara2008p033813,RodriguezLara2009p055806}, and can be used for manipulation of matter at the different scales \cite{Andrews2009}. 

These four propagation invariant families are exact solutions with divergent energies. By using the slowly varying envelope approximation (SVEA), these fundamental families have been shown to support finite energy solutions to Helmholtz equation \cite{GutierrezVega2005p289}; e.~g. Hermite-Gaussian beams are the most known solutions in the Cartesian coordinates. Finite energy beam families are also reported for the cylindrical symmetries, named as Bessel-, Mathieu- and Weber-Gauss beams, which are a closer description for the  optical beams produced in the laboratory, e.~g. those using holographic schemes  \cite{LopezMariscal2006p068001} or optical resonators \cite{AlvarezElizondo2008p18770}. 
A theoretical description close to the experimental schemes of matter manipulation by these families of structured light requires the use of finite energy beams. In practice, the normalization of these structured light gives the requirement to determine the irradiance output needed from the laser sources to produce the desired interchange of mechanical variables. Furthermore, quantization of the corresponding finite energy fields requires the normalization of the scalar wave families. Currently, the general understanding of the parabolic waves, beams and fields includes their propagation characteristics and dynamical variables as well as holographic schemes for beam production   \cite{GutierrezVega2005p289,LopezMariscal2006p068001,VolkeSepulveda2006p867,RodriguezLara2009p055806,Bandres2004p44}. 
But to one's surprise, the normalization for Weber-Gauss beams is missing in the literature. 

In this work, the link between Weber waves and the well-known Bessel and Mathieu waves is provided by deriving the integral relations between parabolic, circular and elliptical waves following the phase space method proposed by Boyer, Kalnis and Miller \cite{Miller1984,Boyer1976p35}. 
With the Bessel and Mathieu wave decompositions of Weber waves, we present the normalization of Weber beams under a series scheme \cite{GutierrezVega2007p215}.
The efficiency to approximate Weber-Gauss beams as a finite superposition of Bessel-Gauss beams is also given.
The results found here should provide the necessary information for the quantization of the corresponding finite energy vector fields and for the understanding of any possible experimental demonstration of mechanical transfer involving such fields.

\section{Weber waves}
The wave equation in parabolic-cylindrical coordinates, $u + \imath~v = \left[ 2 (x + \imath~y)\right]^{(1/2)}$, accepts separable solutions of the form
\begin{eqnarray}
\Psi_{e,k,\gamma,a}^{(W)}(u,v,z)&=& \frac{ \left\vert \Gamma\left[\frac{1}{4} + \frac{\imath a}{2}\right] \right\vert^2}{ \pi \sqrt{2 \sin \gamma}} U_{e,k,\gamma,a}(u) V_{e,k,\gamma,a}(v) e^{\imath ( k_z z - \omega t)},\\
\Psi_{o,k,\gamma,a}^{(W)}(u,v,z)&=& \frac{ \sqrt{2} \left\vert  ~\Gamma\left[\frac{3}{4} + \frac{\imath a}{2}\right] \right\vert^2}{ \pi \sqrt{ \sin \gamma}} U_{o,k,\gamma,a}(u) V_{o,k,\gamma,a}(v) e^{\imath ( k_z z - \omega t)}.
\end{eqnarray}
These expressions, Weber waves, are Dirac delta normalized for a given frequency $\omega$
\begin{equation}
\int d^3\mathbf{x} \Psi_{\tilde{p},k,\tilde{\gamma},\tilde{a}}^{(W)~\ast} \Psi_{p,k,\gamma,a}^{(W)} = \delta_{\tilde{p},p}  \delta(\tilde{\gamma} - \gamma) \delta(\tilde{a} - a).
\end{equation}
The label set $\{p,k,\gamma,a\}$ stands for the parity with respect to coordinate variables $u$ and $v$, even or odd, wave number $k= \omega/c$, Euler angle corresponding to the decomposition of the wave vector in longitudinal and perpendicular components $\mathbf{k} = k_{\perp}~\mathbf{e}_{\perp} + k_{z}~\mathbf{e}_{z}= k \sin \gamma ~\mathbf{e}_{\perp} + k \cos \gamma ~\mathbf{e}_{z}$, and the real continuous eigenvalue, $a$, of the even,
\begin{eqnarray}
U_{e,k,\gamma,a}(u) &=& e^{-\imath k_{\perp} u^2 / 2} ~_1F_1\left(\frac{1}{4} - \frac{\imath a}{2}, \frac{1}{2}, \imath k_{\perp} u^2 \right),\\
V_{e,k,\gamma,a}(v) &=& e^{-\imath k_{\perp} v^2 / 2} ~_1F_1\left(\frac{1}{4} + \frac{\imath a}{2}, \frac{1}{2}, \imath k_{\perp} v^2 \right),
\end{eqnarray}
and odd functions,
\begin{eqnarray}
U_{o,k,\gamma,a}(u) &=& \sqrt{2 k_{\perp}}~ u~ e^{-\imath k_{\perp} u^2 / 2} ~_1F_1\left(\frac{3}{4} - \frac{\imath a}{2}, \frac{3}{2}, \imath k_{\perp} u^2 \right), \\
V_{o,k,\gamma,a}(v) &=& \sqrt{2 k_{\perp}}~ v~  e^{-\imath k_{\perp} v^2 / 2} ~_1F_1\left(\frac{3}{4} + \frac{\imath a}{2}, \frac{3}{2}, \imath k_{\perp} v^2 \right).
\end{eqnarray}
The definition of the hypergeometric $_1F_1$ function follows Ref. \cite{Prudnikov1981v3}. All special functions notation and definitions will follow the latter reference. All calculations are based on the identities presented by Ref. \cite{Prudnikov1981v3,BatemanProject1985v1}.

Weber waves are eigenfunctions for the $z$-component of the linear momentum and the Poisson bracket of the $z$- and $y$-component of the angular and linear momenta, in that order, 
\begin{eqnarray}
P_z \Psi_{p,k,\gamma,a} &=&  k_{z}~ \Psi_{p,k,\gamma,a}, \\
\{J_z, P_y\} \Psi_{p,k,\gamma,a} &=& 2 a k_{\perp}~ \Psi_{p,k,\gamma,a},
\end{eqnarray}
where the notation $\mathbf{P} = -\imath \mathbf{\nabla}$ and $\mathbf{J} = \mathbf{r} \times \mathbf{P}$ has been used for linear and angular momentum. Thus, choosing a positive (negative) real eigenvalue $a$ leads to horizontal parabolas opening to the left (right) with symmetry axis given by the $x$-axis. 

The Plane wave decomposition calculated for these waves is given by 
\begin{eqnarray}
\Psi_{p,k,\gamma,a}(x,y,z) &=&\int_{-\pi}^\pi d\varphi ~\mathfrak{A}^{(W)}_{p,k,\gamma,a}(\varphi)e^{ \imath k_{\perp}(x\cos\varphi + y\sin\varphi)} e^{\imath k_{z} z},
\end{eqnarray}
where the angular spectra is written 
\begin{eqnarray}
\mathfrak{A}_{e,k,\gamma,a}^{(W)}(\varphi) &=& \frac{e^{ia\ln\vert\tan\varphi/2\vert}}{ \sqrt{2 \pi \sin \gamma |\sin \varphi |}},\\
\mathfrak{A}_{o,k,\gamma,a}^{(W)}(\varphi) &=& -\imath  ~\textrm{sgn} \varphi ~\mathfrak{A}_{e,k,\gamma,a}(\varphi).
\end{eqnarray}

Solutions without well defined parity can be constructed such that, 
\begin{eqnarray}
\Psi_{e/o,k,\gamma,a}^{(W)}(r,\phi,z) &=& \frac{1}{\sqrt{2}}[ \Psi_{k,\gamma,a}^{(W)}(r,\phi,z) \pm \Psi_{k,\gamma,-a}^{(W)}(r,\phi,z) ],
\end{eqnarray} 
with angular spectra, 
\begin{eqnarray}
\mathfrak{A}_{k,\gamma,a}^{(W)}(\varphi) &=& \mathfrak{A}_{e,k,\gamma,a}(\varphi) \Theta(\varphi),\\
\mathfrak{A}_{k,\gamma,-a}^{(W)}(\varphi) &=& \imath  ~\mathfrak{A}_{k,\gamma,a}(-\varphi),
\end{eqnarray}
where the notation $\Theta$ represents the Heaviside Theta function. The latter spectra are equivalent to those expressions presented in Ref.\cite{Miller1984,Boyer1976p35}.

\begin{figure}[htp]
$$\begin{array}{ccc}
\includegraphics[width=0.30\textwidth]{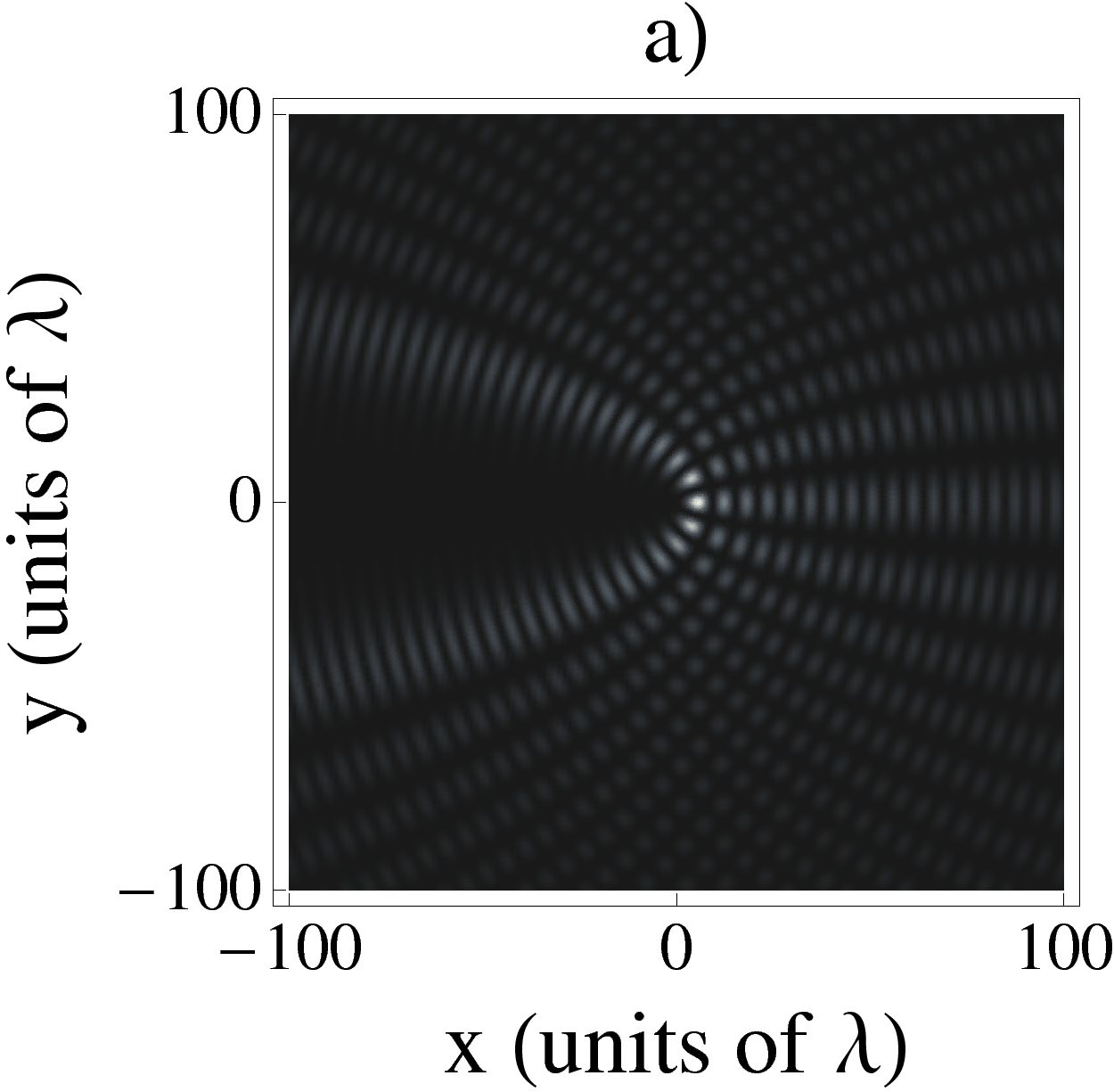}&
\includegraphics[width=0.30\textwidth]{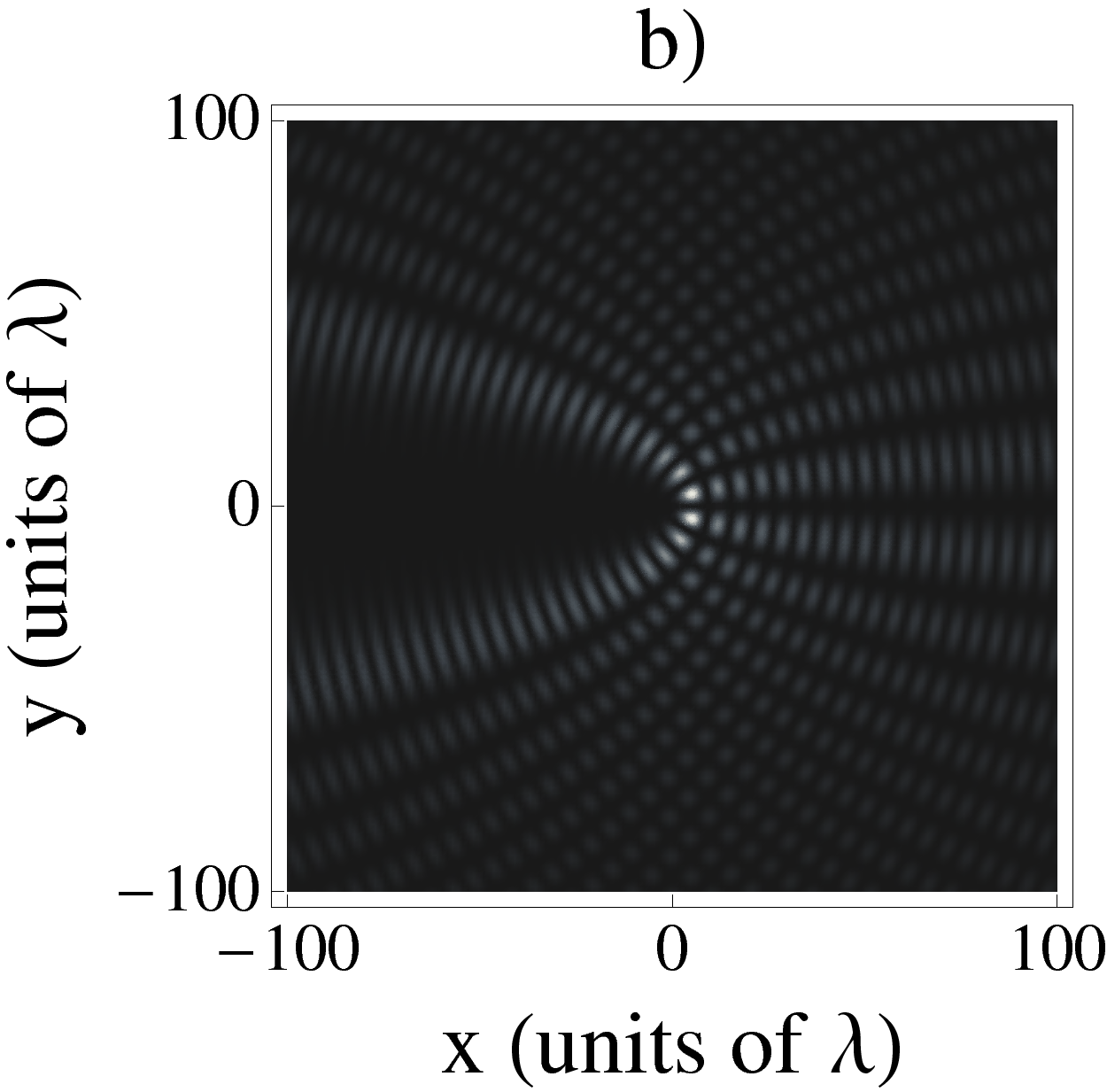}&
\includegraphics[width=0.30\textwidth]{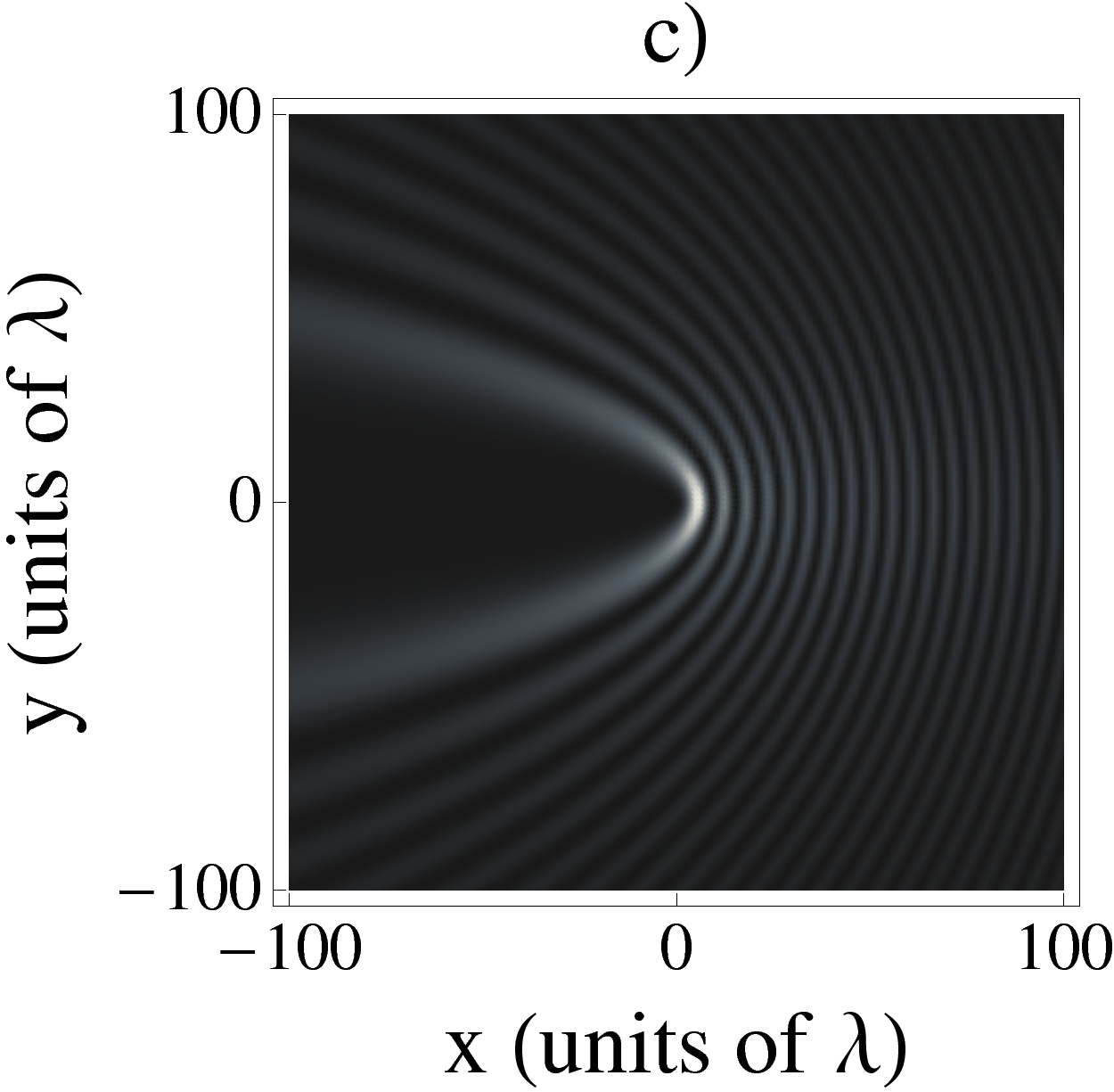}\\
\includegraphics[width=0.30\textwidth]{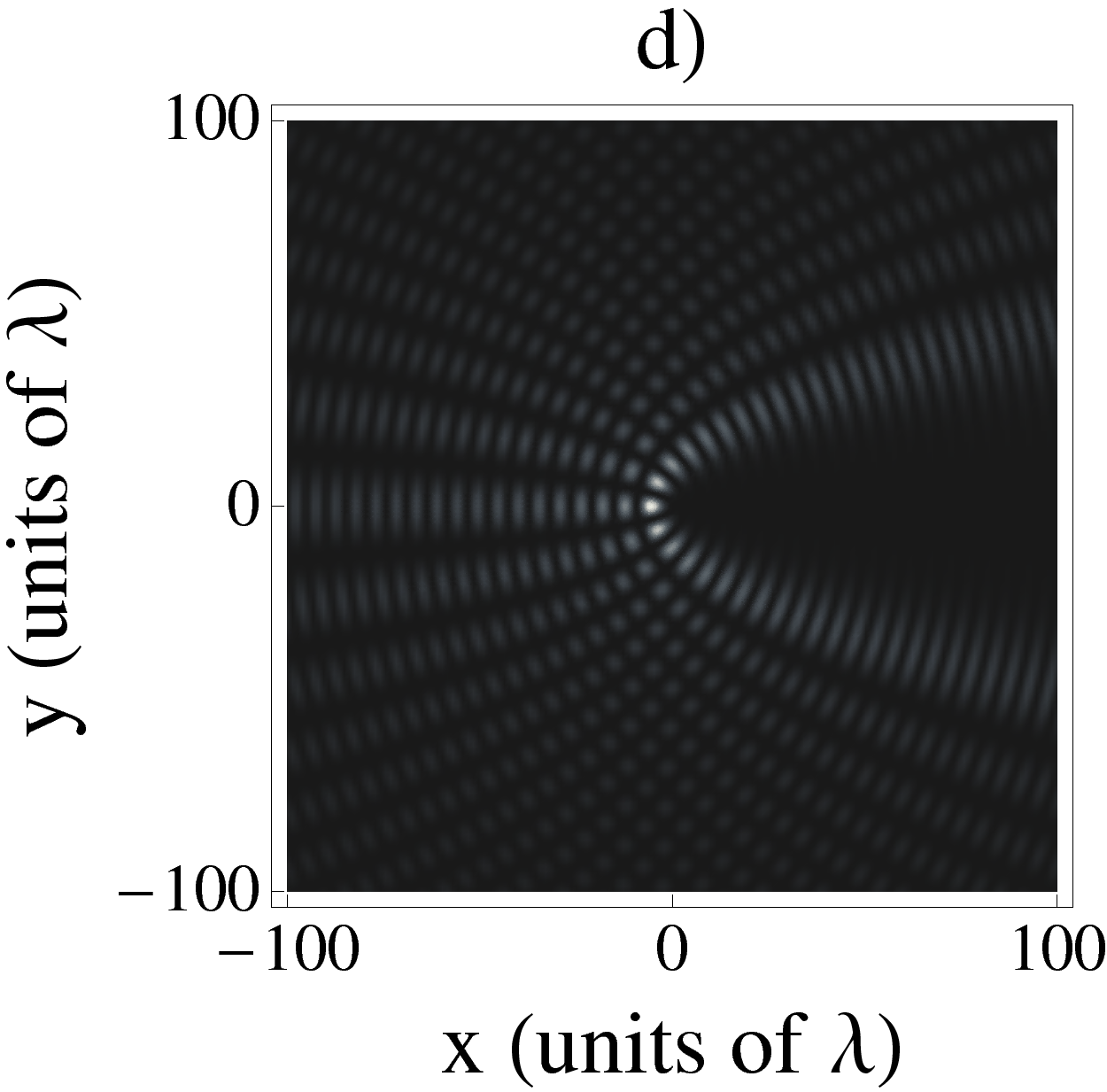}&
\includegraphics[width=0.30\textwidth]{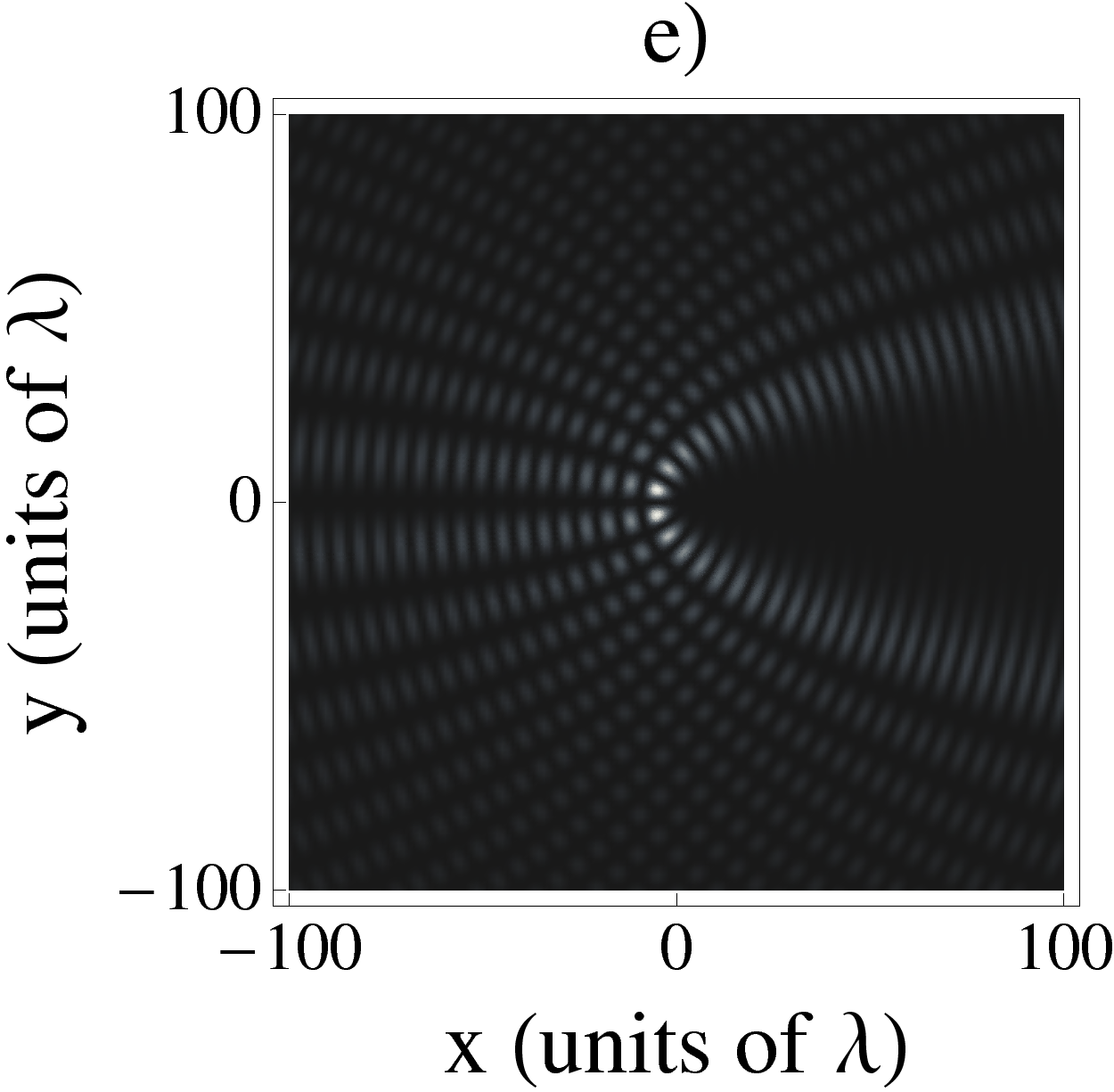}&
\includegraphics[width=0.30\textwidth]{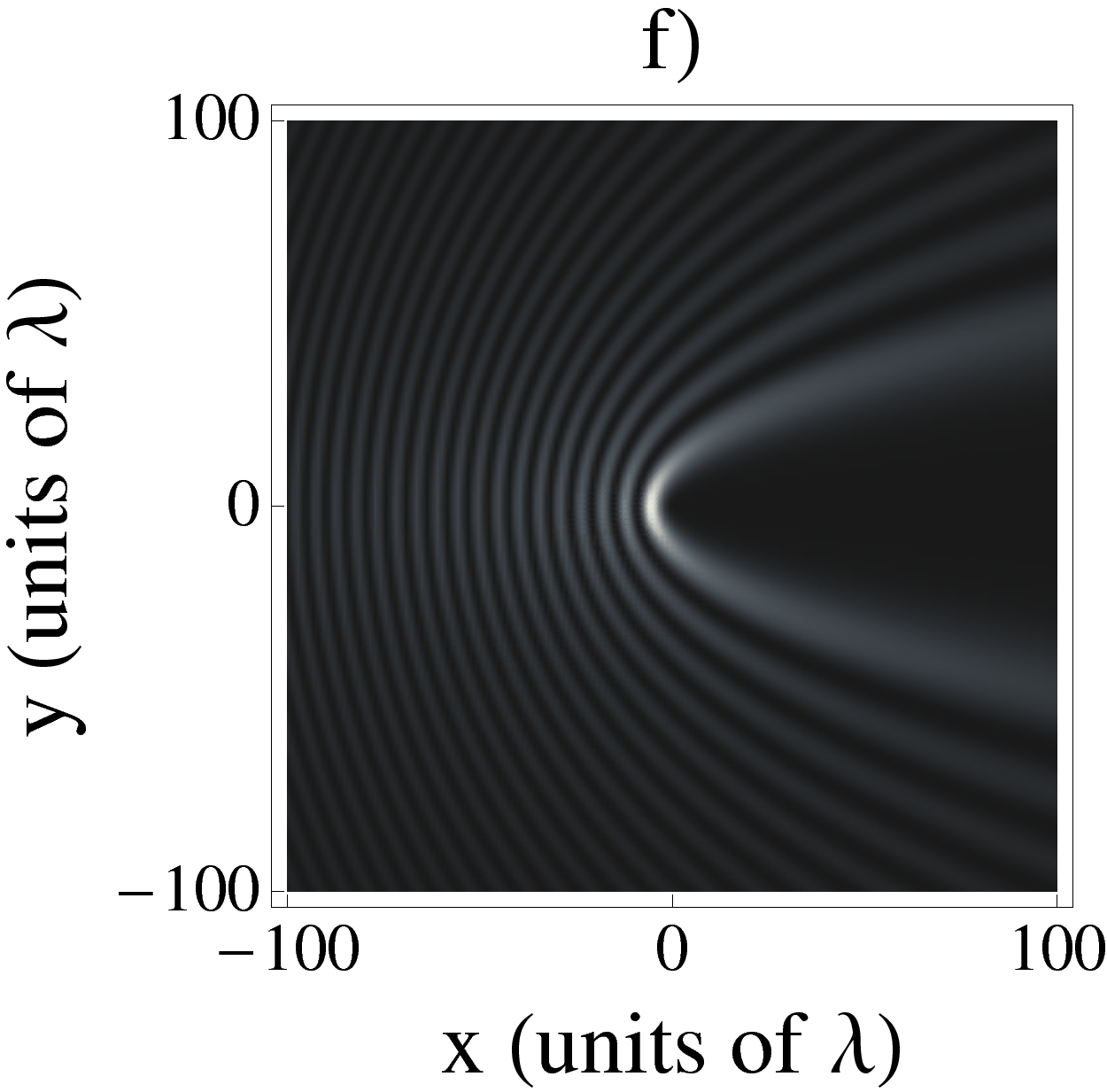}
\end{array}
$$
\caption{Ideal Weber waves with wave number $k = 2 \pi / \lambda$ and $z-$component $\gamma = 0.31756$ rad leading to $k_z= 0.95 ~k$.  a) $\Psi_{e,k,\gamma,a=2}^{(W)} $, b) $ \Psi_{o,k,\gamma,a=2}^{(W)}$, c) $\Psi_{k,\gamma,a=2}^{(W)} $, d) $ \Psi_{e,k,\gamma,a=-2}^{(W)}$, e) $\Psi_{o,k,\gamma,a=-2}^{(W)} $, f) $ \Psi_{k,\gamma,a=-2}^{(W)}$.}
\label{fig:Fig1}
\end{figure}

Figure \ref{fig:Fig1} shows a sampler of Weber waves for a given wave vector as well as positive and negative eigenvalue $a$.

\subsection{Bessel Decomposition}

Using the plane wave decomposition for Weber and Bessel waves, it is possible to show that the Bessel wave decomposition of a Weber wave is given by the superposition
\begin{eqnarray}
\Psi_{p,k,\gamma,a}^{(W)}(r,\phi,z) &=&  \sum_{n =0}^{\infty} \psi_{p,k,\gamma,a}^{(B)}(n) ~\Psi_{p,k,\gamma,n}^{(B)}(r,\phi,z),
\end{eqnarray}
where the coefficients for the expansion are,
\begin{eqnarray} 
\psi_{e,k,\gamma,a}^{(B)}(n) &=& \frac{1}{2 \pi} e^{- \pi (2a +\imath)/4} \left[  C(n,a) + C(-n,a) \right], \\
\psi_{o,k,\gamma,a}^{(B)}(n) &=& -\frac{\imath}{2 \pi} e^{- \pi (2a +\imath)/4} \left[  C(n,a)-C(-n,a) \right], \\
\psi_{k,\gamma,\pm a}^{(B)}(n) &=& \frac{1}{2 \pi} e^{- \pi (2a +\imath)/4} C(\mp n,a),
\end{eqnarray}
with auxiliary functions, 
\begin{eqnarray}
C(n,a) &=&  \Gamma\left(n+\frac{1}{2}\right) \left[ (-1)^{n} f(n,a) + \imath ~\mathrm{sgn}a f(n,-a) \right], \\
f(n,a) &=& \frac{ 2^{\imath a} \Gamma\left(\frac{1}{2} - \imath a \right)}{\Gamma(1 + n - \imath a)}~_2F_1\left( \frac{1}{2} -\imath a, \frac{1}{2} -\imath a,1 + n - \imath a, \frac{1}{2}  \right).
\end{eqnarray}
Notice that the complex conjugate of the auxiliary functions are given by $C^{\ast}(n,a) = C(n,-a)$ and $f^{\ast}(n,a)= f(n,-a)$, while the even and odd Bessel waves are defined in the standard way,
\begin{eqnarray}
\Psi_{e/o,k,\gamma,n}^{(B)}(r,\phi,z) &=& \frac{1}{\sqrt{2}}[ \Psi_{k,\gamma,n}^{(B)}(r,\phi,z) \pm \Psi_{k,\gamma,-n}^{(B)}(r,\phi,z) ], \mbox{ } n=0,1,2,3\ldots
\end{eqnarray} 
as functions of the Dirac delta normalized Bessel wave, 
 \begin{equation}
\Psi_{k,\gamma,n}^{(B)}(r,\phi,z) = (2 \pi)^{(-1/2)}\imath^n J_{n}(k_{\perp} r) e^{\imath (n \phi + k_z ~z)}, \mbox{ }  n = 0, \pm 1, \pm 2, \ldots
\end{equation}
with angular spectra given by the expression
\begin{equation}
\mathfrak{A}_{k,\gamma,n}^{(B)}(\varphi) = \frac{e^{\imath n \varphi}}{ \sqrt{2 \pi \sin \gamma}}.
\end{equation}

\subsection{Mathieu Decomposition}

Eigenfunctions to the wave equation in elliptic-cylindrical coordinates, $x+\imath y = f \cosh[\xi + \imath \eta] $, where $f$ is half the interfocal distance of the coordinate system, are given by Mathieu waves
\begin{eqnarray}
\Psi_{e,k,\gamma,a}^{(M)}(\xi,\eta,z)&=& s_{e,n,q} Ce_{m}(\xi,q) ce_{m}(\eta,q) e^{\imath ( k_z z - \omega t)},\\
\Psi_{o,k,\gamma,a}^{(M)}(\xi,\eta,z)&=& s_{o,n,q} Se_{m}(\xi,q) se_{m}(\eta,q)e^{\imath ( k_z z - \omega t)},
\end{eqnarray}
where the parameter is defined as $q = (f k_{\perp} / 2)^2$ and the normalization coefficients for the four possible families, two parities (even or odd) and two periodicities ($\pi$ or $2\pi$),  are \cite{InayatHussain1991p669}
\begin{eqnarray}
s_{e,2n,q}&=&\frac{\textrm{ce}_{2n}(0,q)\textrm{ce}_{2n}(\pi/2,q)}{A_0^{(2n)}(q)},\\
s_{e,2n+1,q}&=&-\frac{\textrm{ce}_{2n+1}(0,q)\textrm{ce}^\prime_{2n+1}(\pi/2,q)}{q^{1/2}A_1^{(2n+1)}(q)}, \\
s_{o,2n+2,q}&=&\frac{\textrm{se}_{2n+2}^\prime(0,q)\textrm{se}^\prime_{2n+2}(\pi/2,q)}{q B_2^{(2n+2)}(q)},\\
s_{o,2n+1,q}&=&\frac{\textrm{se}^\prime_{2n+1}(0,q) \textrm{se}_{2n+1}(\pi/2,q)}{q^{1/2}B_1^{(2n+1)}(q)}.
\end{eqnarray}
An interesting feature of Mathieu waves is that they take the form of ellipses (hyperbolas) as the relation $a_{m}(q) - 2 q$ or $b_{m}(q) - 2 q$ is positive (negative) with $a_{m}(q)$  the characteristic value for even Mathieu waves and $b_{m}(q)$ for odd.

Mathieu waves have an angular spectra given by the expressions, 
\begin{eqnarray}
\mathfrak{A}_{e,k,\gamma,m}^{(M)}(\varphi) = \frac{\textrm{ce}_{m}(\phi,q)}{ \sqrt{2 \pi \sin \gamma}}, \\
\mathfrak{A}_{o,k,\gamma,m}^{(M)}(\varphi) = \frac{\textrm{se}_{m}(\phi,q)}{ \sqrt{2 \pi \sin \gamma}}.
\end{eqnarray}
As the ordinary even or odd Mathieu functions are defined as a cosine or sine series,
\begin{eqnarray}
\mathrm{ce}_{m} (\phi,q) &=& \sum_{n=0}^{m} A^{(m)}_{\tilde{n}}(q) \cos \tilde{n} \phi, \\
\mathrm{ce}_{m} (\phi,q) &=& \sum_{n=0}^{m} B^{(m)}_{\tilde{n}}(q) \sin \tilde{n} \phi, \quad  \tilde{n} = 2 n + m~\mathrm{mod}~2,
\end{eqnarray}
it is possible to relate the coefficients of the Mathieu decomposition with the aforementioned Bessel decomposition coefficients,
\begin{eqnarray} 
\psi_{e,k,\gamma,a}^{(M)}(m) &=& \sum_{n=0}^{m} A^{(m)}_{\tilde{n}}(q)  \psi_{e,k,\gamma,a}^{(B)}(\tilde{n}),\\
\psi_{o,k,\gamma,a}^{(M)}(m) &=& \sum_{n=0}^{m}  B^{(m)}_{\tilde{n}}(q) \psi_{o,k,\gamma,a}^{(B)}(\tilde{n}), \quad  \tilde{n} = 2 n + m~\mathrm{mod}~2, 
\end{eqnarray}
where the Mathieu decomposition of a Weber wave is given by 
\begin{equation}
\Psi_{p,k,\gamma,a}^{(W)}(\xi,\eta,z) = \sum_{m = 0}^{\infty} \psi_{p,k,\gamma,a}^{(M)}(m)~  \Psi_{p,k,\gamma,m}^{(M)}(\xi,\eta,z) .
\end{equation}

\section{Weber beams}

A Weber beam is given by the SVEA solution to the Helmholtz equation \cite{GutierrezVega2005p289},
\begin{equation}
\Phi_{p,k,\gamma,a}(\mathbf{r}) = \frac{1}{\mu(z)} e^{\imath \left(k -\frac{k_{\perp}^2 }{ 2 k \mu(z)} \right) z} e^{ - \frac{ r_{\perp}^2}{ \omega_{0}^2 \mu(z) }} \Psi_{p,k,\gamma,a} \left(\tilde{u}, \tilde{v} \right).
\end{equation}
The exponential part accounts for the Gaussian envelope with minimum waist $w_0$ and parameter $\mu(z) = 1 + \imath z / z_R$ where the Rayleigh distance is given by $z_R = k \omega_{0}^2 / 2 $ and the modified coordinates are defined as functions of $\tilde{x} = x / \mu(z)$ and $\tilde{y} = y / \mu(z)$. Please notice that the structure of the Weber waves is kept at the $z=0$ plane but it may change with propagation. 

\subsection{Normalization.}
Normalization at the $z=0$ plane yields complex integrals wich can be avoided using a normalization scheme based on the Bessel decomposition \cite{GutierrezVega2007p215}. Under this scheme, normalization for Weber beams yield
\begin{equation}\label{eq:NormSeries}
\int d^2\mathbf{x} |\Phi_{p,k,\gamma,a}(r,\phi)|^2 = \pi \tilde{\omega}^2 e^{- k_{\perp}^2 \tilde{\omega}^2} \sum_{n=0}^{\infty} (1+ \delta_{n,0}) | \psi_{p,k,\gamma,a}^{(B)}(n)|^2 \mathrm{I}_{n} \left( k_{\perp}^2 \tilde{\omega}^2\right),
\end{equation}
where the following notation has been used, $\tilde{\omega} = \omega_0/2$, $\delta_{m,n}$ stands for Kronecker delta, and $I_{\nu}$ for the $\nu$th-order modified Bessel function of the first kind. 

The series is convergent, in general it is possible to numerically argue that 
\begin{equation}
\lim_{n\rightarrow\infty} \frac{S_{n}}{S_{n-1}} < 1,
\end{equation}
for any given value of $a$ with
\begin{equation}
S_{n} = \pi \tilde{\omega}^2 e^{- k_{\perp}^2 \tilde{\omega}^2} (1+ \delta_{n,0}) | \psi_{p,k,\gamma,a}^{(B)}(n)|^2 \mathrm{I}_{n} \left( k_{\perp}^2 \tilde{\omega}^2\right) , \mbox{ } n\ge 0.
\end{equation}
For the special case $a=0$ it is straightforward to show that  
\begin{equation}
\lim_{n\rightarrow\infty} \frac{S_{n}}{S_{n-1}} \approx \frac{\mathrm{I}_{n}\left( k_{\perp}^2 \tilde{\omega}^2\right)}{{I}_{n-2}\left( k_{\perp}^2 \tilde{\omega}^2\right)} \approx \left( \frac{e~ k_{\perp}^2 \tilde{\omega}^2}{8~n} \right)^2, \mbox{ } ~\log e = 1,
\end{equation}
where it has been used,
\begin{eqnarray}
| \psi_{e,k,\gamma,0}(0)|^2 &=& \frac{4 \Gamma^{2}\left( \frac{5}{4}\right)}{\pi \Gamma^{2}\left( \frac{3}{4}\right)},\\
| \psi_{e,k,\gamma,0}(n)|^2 &=& \left( 1 - n ~\mathrm{mod}~ 2  \right) \frac{ \Gamma^{2}\left( \frac{2 n + 1}{4}\right)}{4 \pi \Gamma^{2}\left( \frac{2 n + 3}{4}\right)}, \mbox{ } n=1,2,3,\ldots\\
| \psi_{o,k,\gamma,0}(1)|^2 &=& \frac{4 \Gamma^{2}\left( \frac{3}{4}\right)}{\pi \Gamma^{2}\left( \frac{1}{4}\right)},\\
| \psi_{e,k,\gamma,0}(n)|^2 &=& \left( n ~\mathrm{mod}~ 2 \right) \frac{ \Gamma^{2}\left( \frac{2 n + 1}{4}\right)}{4 \pi \Gamma^{2}\left( \frac{2 n + 3}{4}\right)}, \mbox{ } n=2,3,4,\ldots
\end{eqnarray}
and the asymptotic limits \cite{Lebedev1965}
\begin{eqnarray}
\mathrm{I}_{n}(z) \approx \frac{1}{\sqrt{2 \imath}} e^{\frac{\imath n \pi}{2} + n + n \log(\frac{z e^{- \imath \pi /2}}{2}) - (n+\frac{1}{2}) \log n }, \quad n \rightarrow \infty, \\
\Gamma(z) \approx \sqrt{2 \pi} e^{-z} z^{z -1/2} \left(1 + \frac{1}{12 z} \right), \quad z \rightarrow \infty.
\end{eqnarray}

\begin{figure}[htp]
\begin{center}
\includegraphics[width=0.45\textwidth]{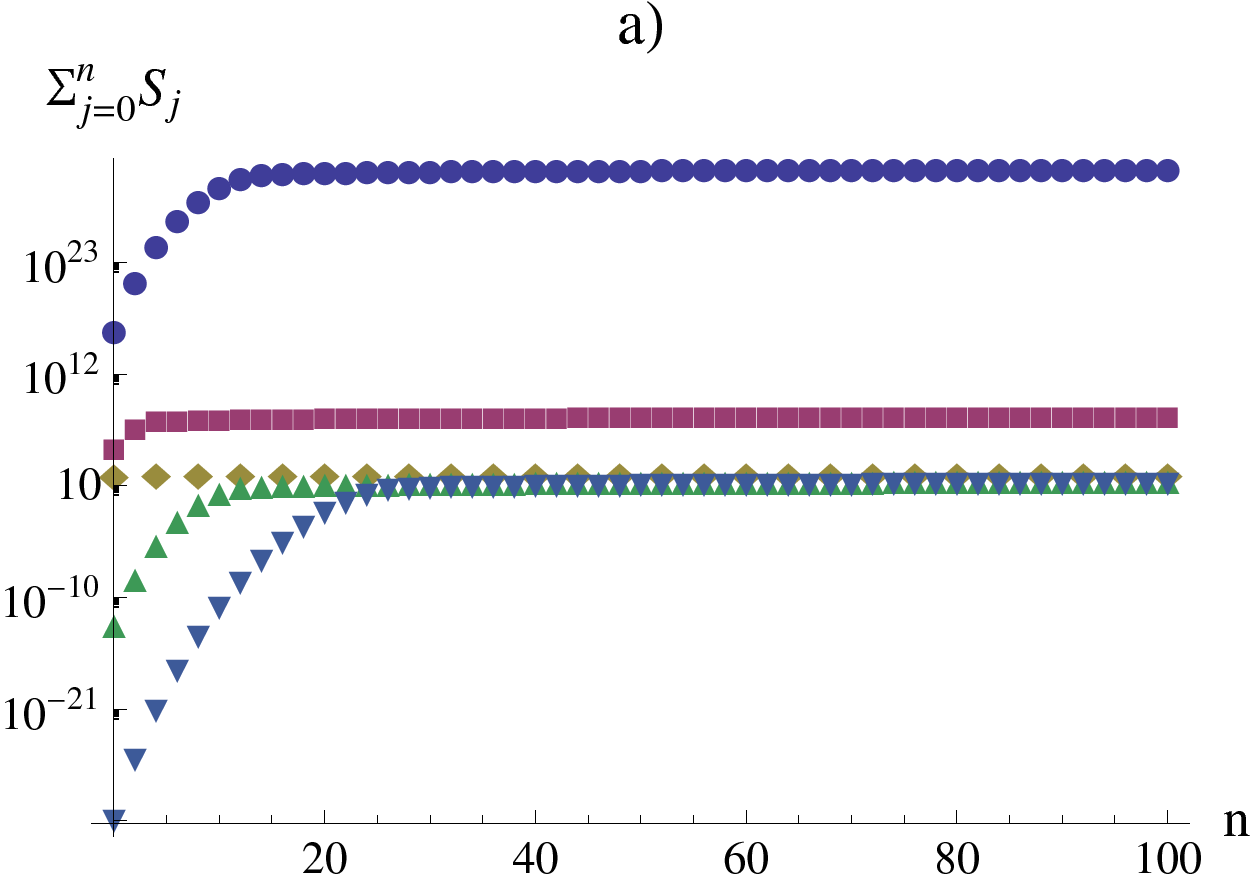}\\
\includegraphics[width=0.45\textwidth]{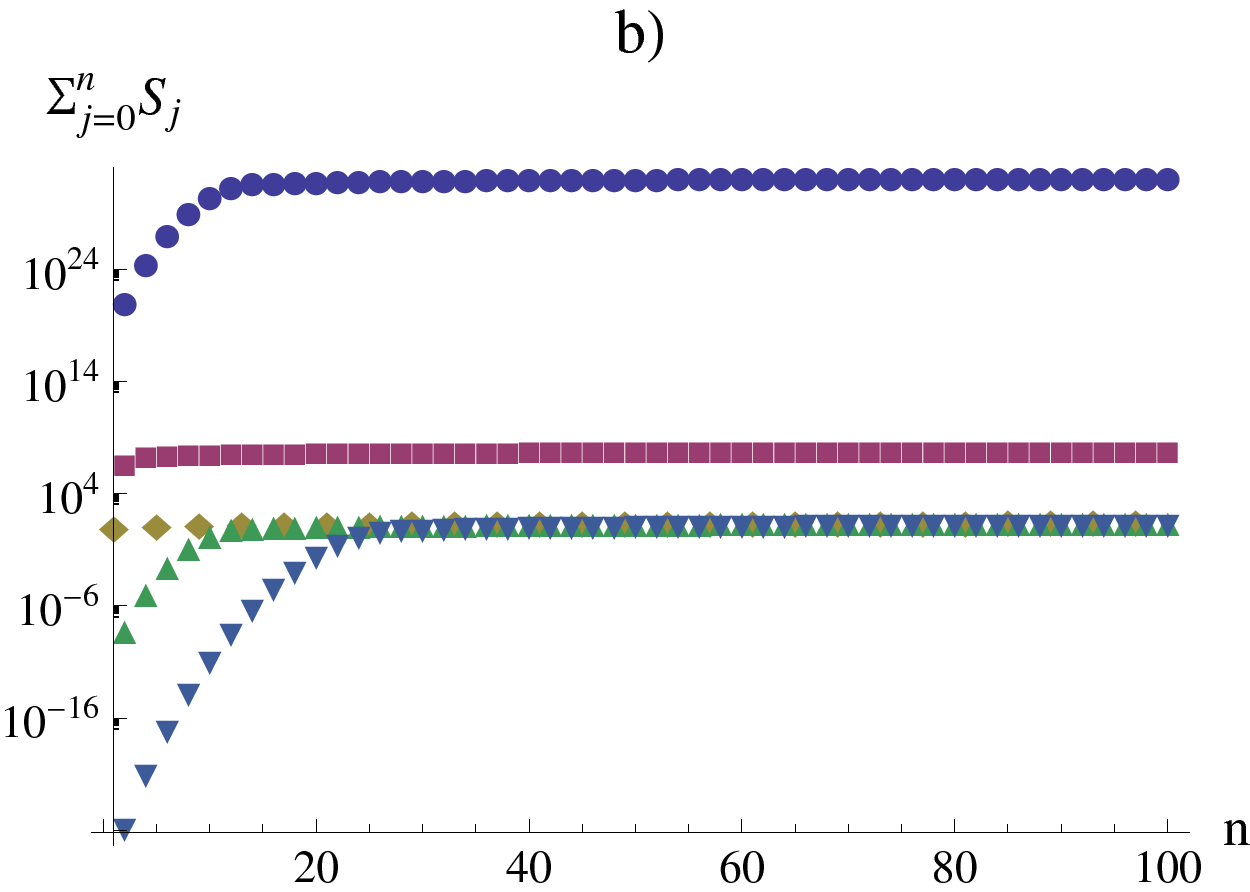}
\caption{(Color online) Behaviour of the truncated normalization coefficient $\sum_{j=0}^{n} S_{j}$ for a given wave number $k = 2 \pi / \lambda$ and Euler angle $\gamma = 0.31756$ rad leading to $k_z= 0.95 ~k$ for a) Even Weber fields $\Psi_{e,k,\gamma,a}^{(W)} $ and b) $ \Psi_{o,k,\gamma,a=2}^{(W)}$ with eigenvalues $a=-11.2821$ (circle), $a=-2.2827$ (square), $a=0$ (diamond), $a=10$ (triangle), $a= 23.7902$ (upside-down triangle).}
\label{fig:Fig2}
\end{center}
\end{figure}

Figure \ref{fig:Fig2} shows the behaviour of the truncated normalization coefficient, $\sum_{j=0}^{n} S_{j}$, as a function of the total number of accounted terms, $n$, for even and odd Weber beams for a given wave vector and some random values of $a$. A thorough sampling in the range $a \in [-25,25]$ showed that the truncated normalization coefficient was well stabilized at around the fiftieth term, at most. 

\subsection{Generation.}

The holographic generation of a Weber wave or beam implies a great loss of irrandiance. An optical resonator with parabolic modes as output has not been reported yet to the knowledge of the author. Thus, a feasible scheme for obtaining a high irradiance Weber beam relies on the superposition of Bessel or Mathieu beams from optical resonators \cite{AlvarezElizondo2008p18770}. 

In order to realize how many Bessel or Mathieu beams are required to reproduce a Weber beam, the behaviour of the coefficients for the Bessel or Mathieu decomposition has to be studied. Although the asymptotic limit for the modified Bessel function of the first kind, the Hypergeometric and the Gamma function are known \cite{Lebedev1965, Abramowitz1970}, it is quite complex to get an analytical asymptotic behaviour for the ratio between Bessel or Mathieu decomposition coefficients. 

\begin{figure}[htp]
\begin{center}
\includegraphics[width=0.45\textwidth]{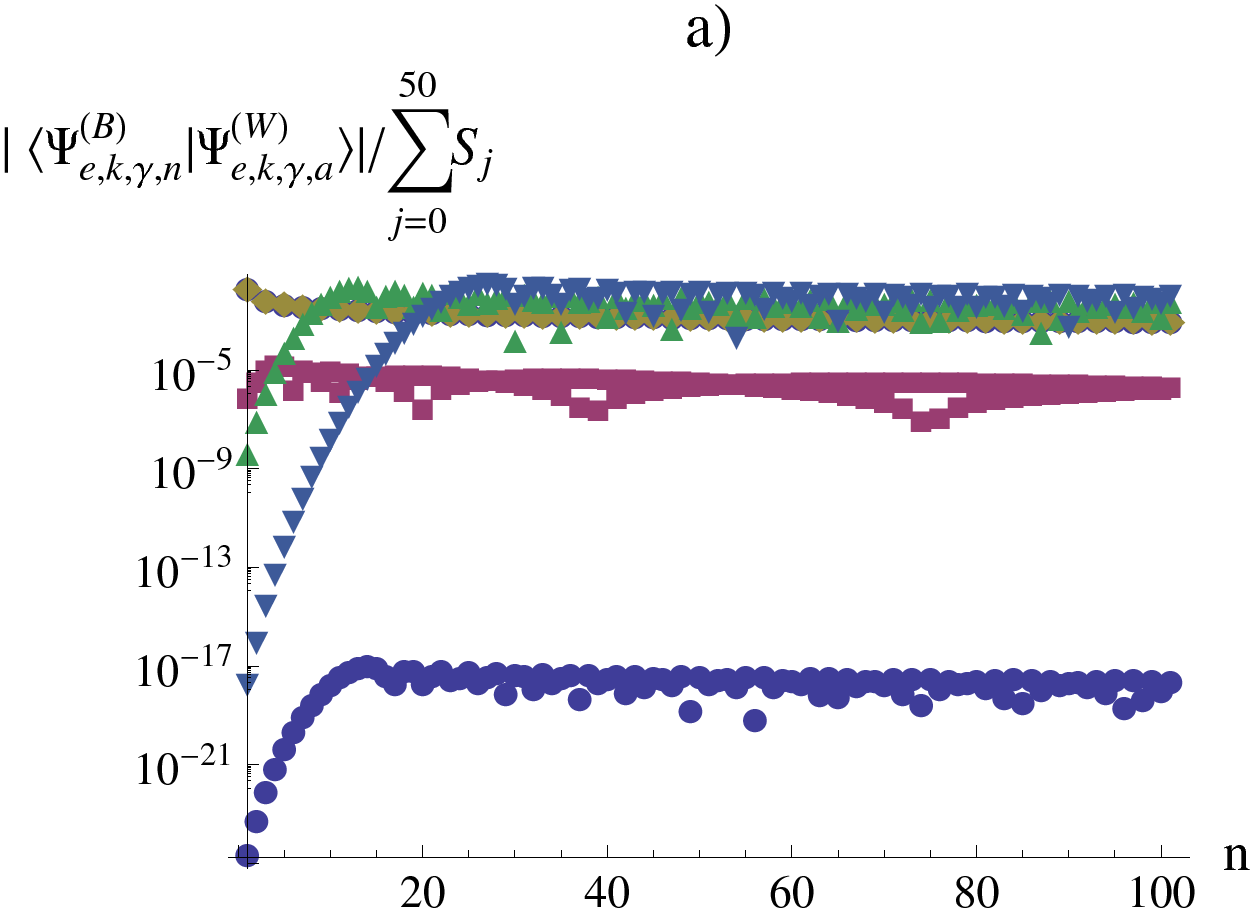}\\
\includegraphics[width=0.45\textwidth]{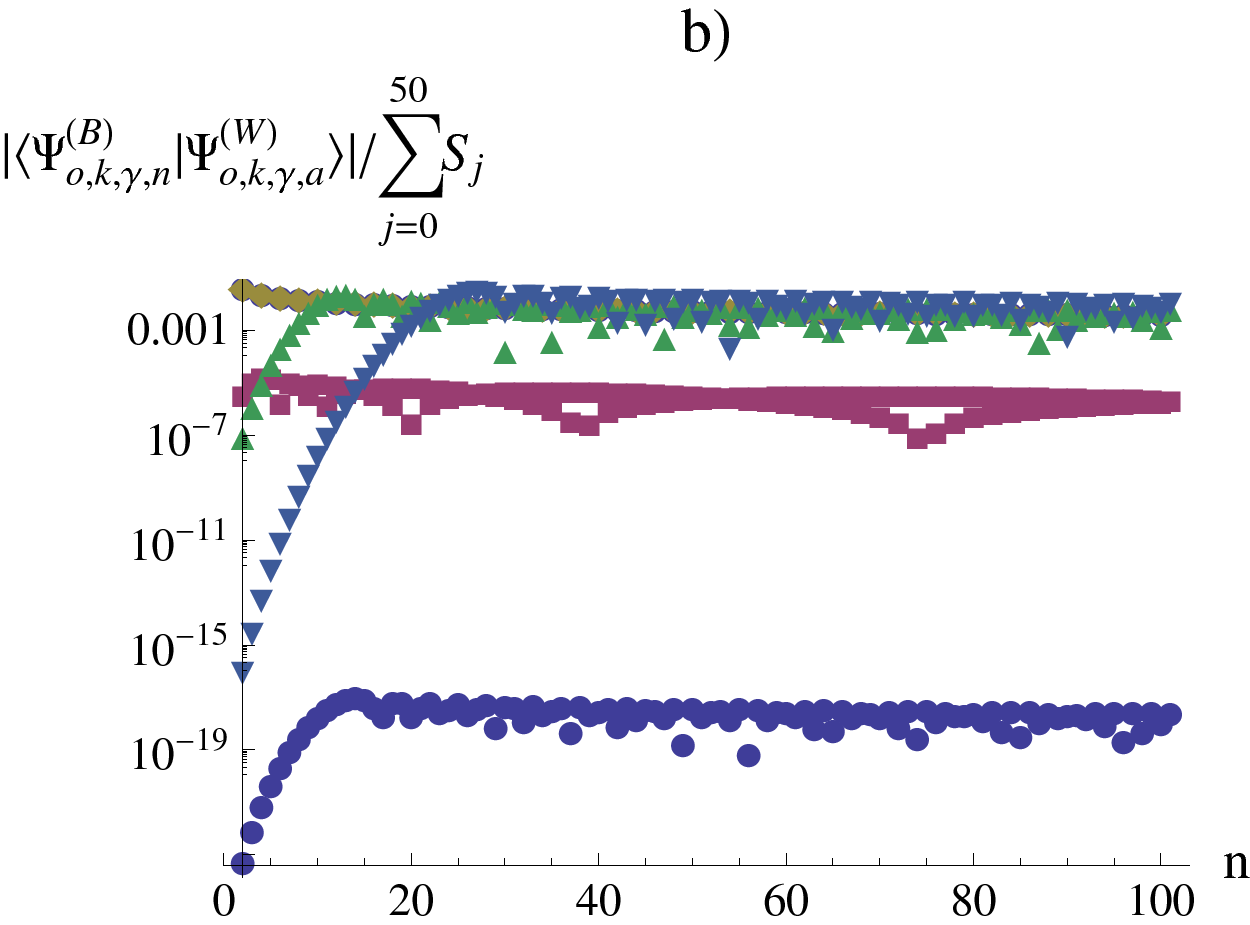}
\caption{(Color online) Behaviour for the absolute value of the normalized Bessel decomposition coefficients of a) even  and b) odd Weber waves with a given wave number $k = 2 \pi / \lambda$, Euler angle $\gamma = 0.31756$ rad leading to $k_z= 0.95 ~k$, and eigenvalues $a=-11.2821$ (circle), $a=-2.2827$ (square), $a=0$ (diamond), $a=10$ (triangle), $a= 23.7902$ (upside-down triangle).}
\label{fig:Fig3}
\end{center}
\end{figure}

For the Bessel decomposition, a thorough numerical survey for large values of $n$ in the range $a\in [-25,25]$ shows that  
\begin{equation}{\label{eq:LimitBesselD}}
\lim_{n\rightarrow\infty} \left\vert \frac{\psi_{p,k,\gamma,a}^{(B)}(n)}{\psi_{p,k,\gamma,a}^{(B)}(n-1)} \right\vert \rightarrow 1.
\end{equation} 
This can only be analytically asserted for $a=0$. Figure \ref{fig:Fig3} shows the numerical behaviour for some Bessel decomposition coefficients. It is possible to see that the main contributing terms are not the first terms of the series, with the exception of $a=0$, as $n$ increases the contribution of each term becomes important until it reaches a maximum,  then it oscillates to stabilize and fulfil Eq.\eqref{eq:LimitBesselD}.

\section{Discussion}

The Dirac delta normalization of the even and odd ideal scalar parabolic waves given as hypergeometric functions, Weber waves, has been presented. The integral relations between these eigenfunctions of the Helmholtz equation with parabolic-cylindrical symmetry and those with circular and elliptical-cylindrical symmetries, Bessel and Mathieu waves,  have been shown and used to introduce the Bessel and Mathieu wave decomposition of Weber waves. 

A normalization for the finite energy Weber-Gauss beams was presented based on their Bessel-Gauss decomposition. It has been shown that it is not feasible to efficiently construct a Weber-Gauss beam through the finite superposition of just a few Bessel-Gauss beams.

Finding a close analytical form for the normalization integral straight from the configuration or phase space representation of a Weber beam and the analysis pertaining the generation of Weber beams as the superposition of Mathieu beams is left as an open problem due to the complexity of the calculations involved.

\acknowledgements
The author is grateful to Prof. W. Miller for pointing to helpful references and thanks Prof.  R.~K. Lee, Prof. R. J{\'a}uregui and A. Stoffel for their useful comments. This work was supported by the National Tsing-Hua Univesity under contract No. 98N2309E1.

\end{document}